\documentclass[prx,singlecolumn]{revtex4-2}
\usepackage[utf8]{inputenc}
\usepackage{graphicx}
\usepackage{amsfonts}

\usepackage{titlesec}
\usepackage{tcolorbox}
\usepackage{geometry}
\titleformat{\section}
  {\normalfont\fontsize{11}{12}\bfseries}{\thesection}{1em}{}
\titleformat{\subsection}
  {\normalfont\fontsize{10}{12}\bfseries}{\thesubsection}{1em}{}

\newcommand{\fs}[1]{{#1}} 

\begin{document}
\title{Tackling the subsampling problem to infer collective properties from limited data}

\affiliation{}
\author{Anna Levina$^{1,2}$}
\author{Viola Priesemann$^{3,4}$}
\author{Johannes Zierenberg$^{3}$}

\affiliation{
  $^1$\mbox{Eberhard Karls University of Tübingen, Maria-von-Linden 6, 72076 T{\"u}bingen, Germany}
  $^2$\mbox{Max Planck Institute for Biological Cybernetics, Max-Planck-Ring 8,  72076 T{\"u}bingen, Germany}
  $^3$\mbox{Max Planck Institute for Dynamics and Self-Organization, Am Fa\ss berg 17, 37077 G{\"o}ttingen, Germany},\\
  $^4$\mbox{Institute for the Dynamics of Complex Systems, University of G\"ottingen, Friedrich-Hund-Platz 1, 37077 G\"ottingen, Germany},\\
}

\maketitle 

\section*{Abstract} 
Complex systems are fascinating because their rich macroscopic properties emerge from the interaction of many simple parts.  Understanding the building principles of these emergent phenomena in nature requires assessing natural complex systems experimentally. 
However, despite the development of large-scale data-acquisition techniques, experimental observations are often limited to a tiny fraction of the system. 
This \textit{spatial subsampling} is particularly severe in neuroscience, where only a tiny fraction of millions or even billions of neurons can be individually recorded. 
Spatial subsampling may lead to significant systematic biases when inferring the collective properties of the entire system naively from a subsampled part.
To overcome such biases, powerful mathematical tools have been developed in the past. 
In this perspective, we overview some issues arising from subsampling and review recently developed approaches to tackle the subsampling problem. 
These approaches enable one to assess, e.g., graph structures, collective dynamics of animals, neural network activity, or the spread of disease correctly from observing only a tiny fraction of the system.
However, our current approaches are still far from having solved the subsampling problem in general, and hence we conclude by outlining what we believe are the main open challenges.
Solving these challenges alongside the development of large-scale recording techniques will enable further fundamental insights into the working of complex and living systems. 


\section*{Introduction} 

%
\fs{Complex systems, and in particular living systems, are composed of a multitude of units with diverse interactions: particles are colliding, cells are growing, moving and dividing, agents are communicating, and neurons are sending signals over long distances.}
These interactions give rise to a range of emergent collective phenomena~\cite{munoz_colloquium_2018}, which include pattern formation~\cite{cross_pattern_1993, gollub_pattern_1999,cross_pattern_2009, desai_dynamics_2009, halatek_rethinking_2018}, synchronization~\cite{acebron_kuramoto_2005,arenas_synchronization_2008}, flocking~\cite{cavagna_physics_2018, rahmani_topological_2021}, percolation~\cite{christensen_complexity_2005, araujo_recent_2014, korchinski_criticality_2021}, and self-organized criticality~\cite{bak_self-organized_1987, jensen_self-organized_1998, dickman_paths_2000,aschwanden_self-organized_2011,pruessner_self-organised_2012, hesse_self-organized_2014, zeraati_self-organization_2021}.
Understanding such collective phenomena and their origin is necessary to understand, predict and influence complex systems~\cite{boccaletti_complex_2006}. 

However, most complex systems of interest are comprised of so many units that one cannot observe all of them with full resolution. Therefore, in experiments one needs to rely either on \textit{coarse sampling}, thus indirect measures of their collective properties - with the advantage to potentially cover the full system; or one reverts to \textit{spatial subsampling}  (see Box 1 for a formal definition), by focusing on a precise observations of a typically very small subset of the system. In both cases, the choice of the sampling Ansatz can strongly impact and even bias the observables and hence challenges the understanding of the collective properties~\cite{stumpf_subnets_2005, serafino_true_2021, morstatter_is_2013, priesemann_subsampling_2009,  magni_visualization_2009, chen_avalanche_2011}.

When deciding to sample the full system - but at coarse resolution - one basically assesses the emergent large-scale properties of the system directly.
However, such coarse, indirect measures provide only macroscopic ``observables'' and thereby might fall short of achieving sufficient insight into the microscopic interactions between units~\cite{zierenberg_description_2020,neto_unified_2020} that are essential to explain emergent complex phenomena. 
It is then the great challenge to fill in the missing microscopic parameters using temporally and spatially precise measurements that are, however, only available as severely subsampled observations.  
To advance our understanding of complex systems, we need to understand the challenges that come with the limited sampling, and then bridge the gap between the sampled and the full system to achieve a complete understanding of the emergent phenomena. 

\fs{For systems comprised of a huge number of interacting units, a high resolution sampling of individual components typically comes at the cost of an incomplete sample.}
This problem of spatial subsampling is particularly severe when studying time-evolving systems, for which it is not sufficient to gather data sequentially, even if this way we could observe the whole system at different time-steps.
For example, when studying disease spread a significant fraction of infections might be unreported~\cite{zhao_estimating_2020}, or when studying collective animal dynamics in the field one has limited tracking devices, e.g., to study navigation~\cite{berdahl_collective_2018} or ecological interactions~\cite{de_aguiar_revealing_2019}.
Additionally, even the observations of individual components sometimes cannot be easily obtained, e.g., recording of single neuron typically implies some distortion or modification of brain tissue or cells, and observation of animals in the wild or of social systems may affect their behavior. 
In general, stronger perturbations are expected when more of the units are observed, and often there is a natural upper limit on how many of the units can be observed. 
Recording neurons with electrodes in the brain, for example, is strongly limited by the space required for each electrode, such that only a tiny fraction of the millions or billions of neurons can be individually recorded~\cite{priesemann_subsampling_2009,levina_subsampling_2017,priesemann_spike_2014,ribeiro_undersampled_2014,priesemann_neuronal_2013}.
Typically, one thus has to rely on simultaneous observations of a subset of individual units.
Such spatial subsampling can severely bias the inference about the systems' collective properties.

\begin{tcolorbox}
\textbf{Box 1: Definition of spatial subsampling}\newline
\begin{minipage}[t]{1\linewidth}
\begin{centering}
     \includegraphics[width = 0.6\linewidth]{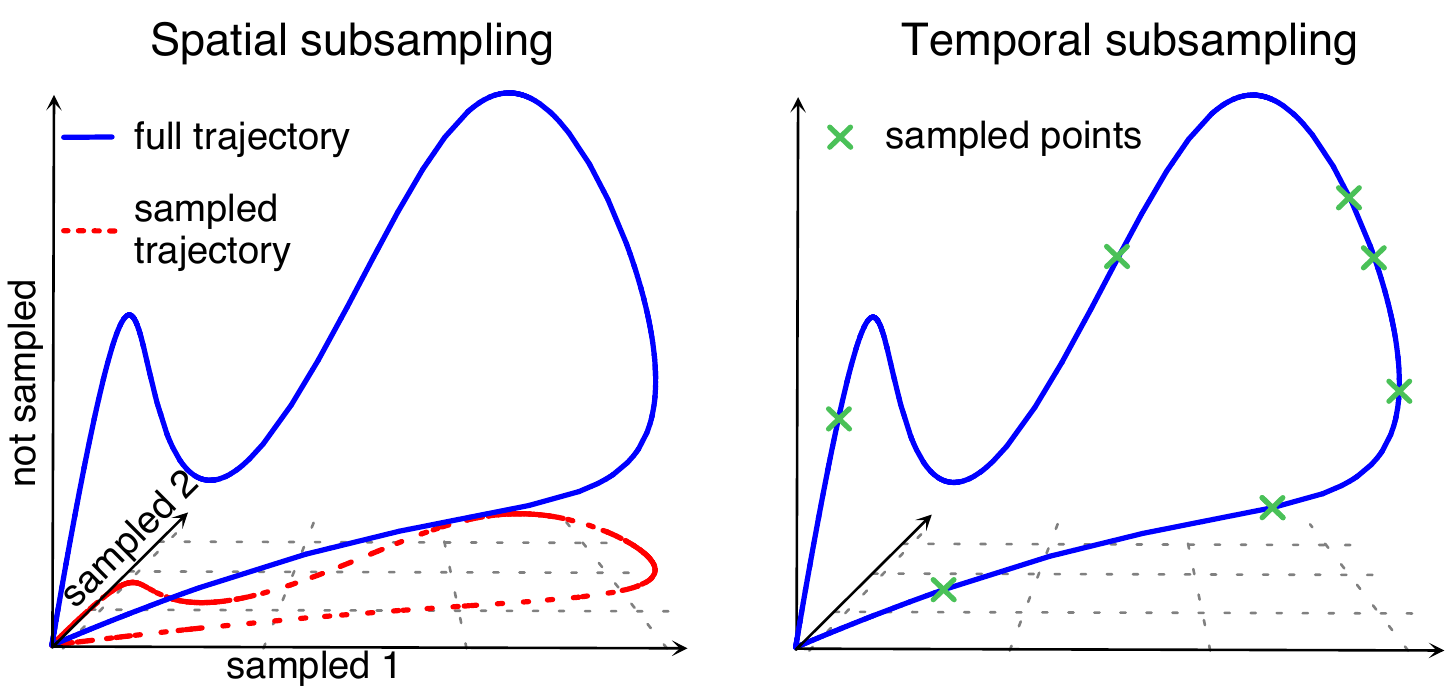}
    \label{fig:math_subsampling}
\end{centering}
\end{minipage}
The problem of spatial subsampling is that of having access only to a fraction of a high-dimensional system (Fig.~top, left). 
This is distinct from the problem of measuring a complete system at only a discrete subset of times (Fig.~top, right), which is a common problem addressed in signal processing.
To formalize what we refer to under spatial subsampling, let us assume a full system of $M$ components with trajectory  $\mathbf{x}^M(t)=(x_1(t),\ldots,x_M(t)), \: t\in[0,T]$ that is generated from a stationary $M \times T$ - dimensional distribution $P^M(\mathbf{x}^M)$. \emph{Samples} from this distribution give the momentary realization of the system, in a \emph{subsample} we observe a subset of all dimensions, i.e. for $N \ll M$ we have access to $\mathbf{x}^N_{\mathrm{obs}}(t)=(x_{i_1}(t),\ldots,x_{i_N}(t))$ that can be described by an $N\times T$-dimensional distribution $Q^N(\mathbf{x}^N_{\mathrm{obs}})$. 
The relation between $P$ and $Q$ is not necessarily straightforward and often we can access it only using a low-dimensional summary statistics $\xi$. The aim thus is to derive a general mapping between $P$ and $Q$ (or at least $\xi(P)$ and $\xi(Q)$).
The desired mapping will, however, strongly depend on the type of subsampling, examples of which we illustrate in the Figure below (left, middle) in comparison to temporal subsampling (right)
\begin{minipage}[t]{1\linewidth}
\begin{centering}
     \includegraphics[trim=0 0 0 0, clip, width = 0.85\linewidth]{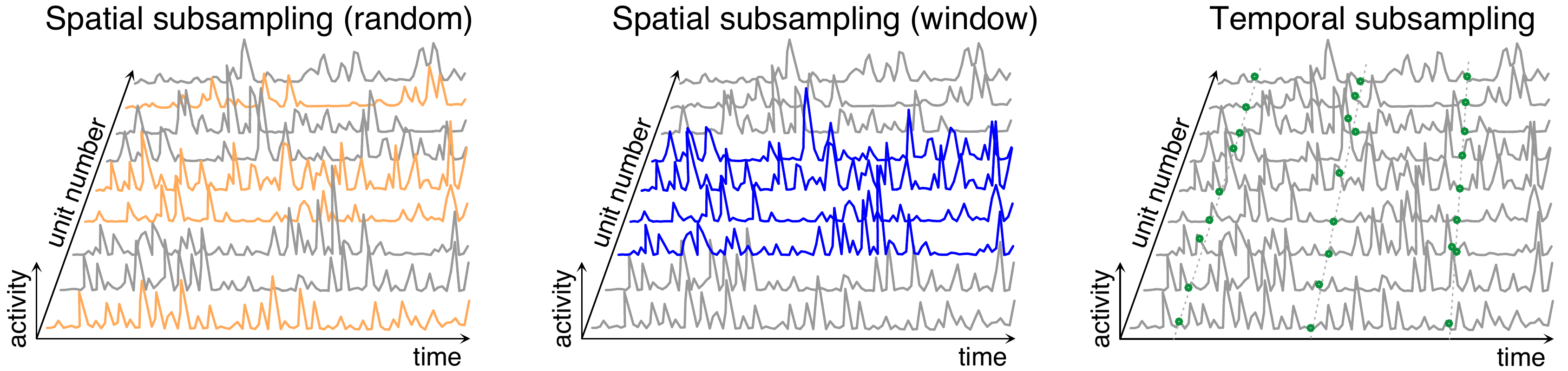} 
    \label{fig:math_subsampling2}
\end{centering}
\end{minipage}
\end{tcolorbox}

\fs{Spatial subsampling can lead to systematic misestimations of collective properties, which typically cannot be overcome by optimizing the sampling process.}
To start with a tutorial illustration, lets recall that for short recordings of a correlated process, estimates of the variance can be biased (see Box 2).
Similarly, when observing only a few units from a large system, the variance of the estimate is obviously decreasing with longer recordings, however, these estimates themselves may still be systematically biased~\cite{wilting_inferring_2018}.
An appropriate choice of the spatial sampling approach, however, can make inference about the full system feasible.
Here, as in most statistical estimates, the construction of a representative sample is crucial, e.g. via representative surveys that may serve as empirical data for sociophysics models~\cite{sznajd-weron_sznajd_2005, castellano_statistical_2009, dinkelberg_detect_2021}.
However, the core challenge that we face for spatial subsampling, is that forming such representative sampling is not always  possible, and even when using a sufficiently representative sampling, the inference about the full system is often still not trivial.
\begin{tcolorbox}
\textbf{Box 2: Incomplete sampling can lead to biased estimation}\newline
    \begin{centering} 
    \includegraphics[width = 0.85\linewidth]{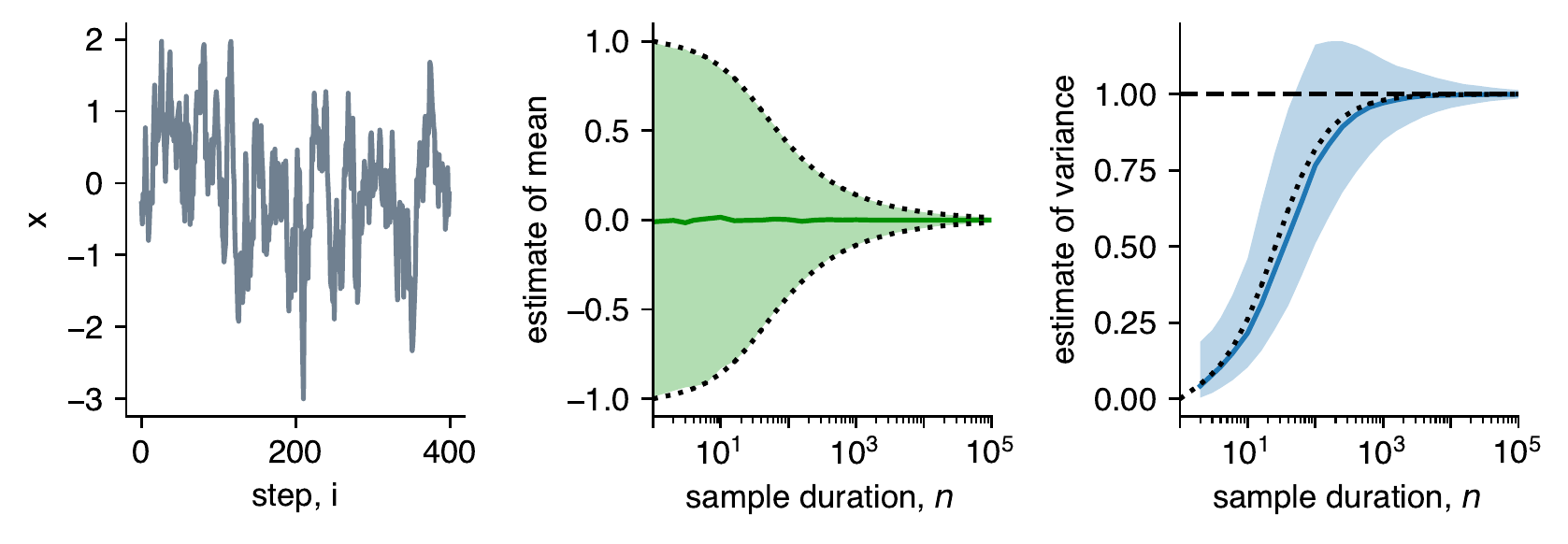} 
    \end{centering}\\
    It is long known that finite data can lead to severe biases when estimating covariances of correlated variables~\cite{marriott_bias_1954}. 
    To illustrate this, we construct each sample as a series of correlated Gaussian variables of finite length $n$ , $\{x_1,x_2,...x_i,...,x_n\}$, with zero mean $\langle x\rangle=0$, unit variance $\sigma^2=\langle x^2\rangle=1$ and non-vanishing (temporal) correlation $\langle x_{i}x_{i+k}\rangle=e^{-k/\tau}$ with $\tau=10$.
    Practically, we construct a bivariate time series~\cite{grotendorst_statistical_2002}  $x_i = \rho x_{i-1} + \sqrt{1-\rho^2}\eta_i$, where $\eta_i$ is drawn independently from a Gaussian distribution with zero mean and unit variance. 
    Explicitly, $x_k=\rho^kx_0 +\sqrt{1-\rho^2}\sum_{l=1}^k\rho^{k-l}\eta_l$ such that $\langle x_k x_0\rangle=\rho^k=e^{-k/\tau}$. An example of such a time series is shown on the left.
    
    To demonstrate the effect of a finite observation duration (i.e. finite number of observables), we focus on estimates of mean and variance from $10^4$ realizations (each a time series of duration $n$).
    Specifically, for each realization we estimate the mean $\overline{x}=\frac{1}{n}\sum_{i=1}^n x_i$ and variance $\hat{\sigma}^2=\frac{1}{n-1}\sum_{i=1}^n (x_i-\overline{x})^2$.
    As expected, we find that the mean computed from individual samples scatters broadly around the asymptotic expectation value $\langle x\rangle = 0$ with the interquartile range (shaded green area) decreasing with $n$, showing that the estimate $\overline{x}$ of the mean has a large variance but is unbiased. 
    In contrast, the estimate of the variance is biased. For small $n$ it clearly deviates from the true variance $\sigma^2 = 1$ (right panel). 
    The reason for this bias is easy to see from the example trace: Since excursions are correlated, the fluctuations about the \textit{local} mean are much smaller than the true variance. Doted lines represent analytical results.
\end{tcolorbox}

\fs{To overcome biases from spatial subsampling, the development of systematic mathematical approaches is required.}
Even when the sampling is in principle representative, na\"ive extrapolations can incur severe biases~\cite{stumpf_subnets_2005, priesemann_subsampling_2009, serafino_true_2021}. 
The bias depends on the system properties, the sampling approach, and the observable of interest. 
Dismissing this bias without proper consideration can lead to contradictory statements, e.g., about the prevalence of the scale-free networks in nature~\cite{serafino_true_2021,voitalov_scale-free_2019, stumpf_critical_2012, broido_scale-free_2019,stumpf_subnets_2005}, or about the dynamical state of the brain~\cite{priesemann_subsampling_2009,priesemann_spike_2014,ribeiro_undersampled_2014,levina_subsampling_2017,neto_unified_2020}. 
To date, we do not have a unifying theory to overcome the subsampling challenge in general. 
However, for several fairly generic settings, systematic mathematical and empirical solutions have been developed in recent years.
In this perspective, we summarize recently developed approaches to infer the collective properties of the entire system from a spatially subsampled observation.
We discuss recent applications mainly using examples from neuroscience and from scale-free networks, where subsampling is a particularly strong hurdle, and hence, many of these techniques originate; however, these methods are much more general and have potential applications to morphogenesis in living systems, to the spread of a disease, to news-propagation in social networks, and beyond. 


\begin{figure}
    \centering
     \includegraphics[trim=0 0 0 0, clip, width = 0.60\linewidth]{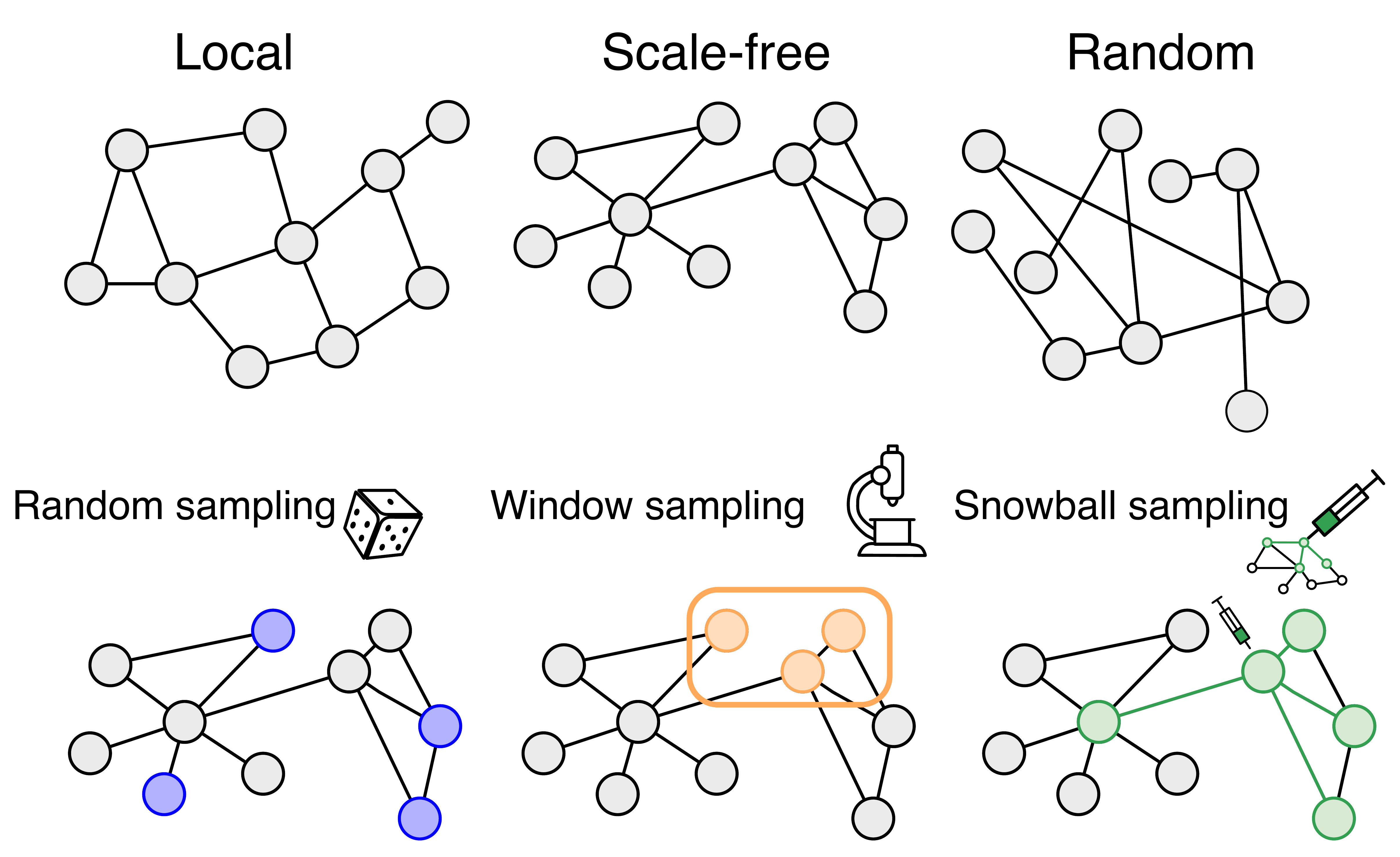}
    \caption{\textbf{Different classes of subsampling}. 
     Top: Sketches of typical structures of spatially embedded networks. Bottom: different approaches of sampling from a spatially-extended system - illustrated with the example of a scale-free network. 
     Random sampling has the advantage of drawing a representative set of nodes. Windowed sampling provides a good local resolution, but the sample might not be representative for the full system. Snowball sampling follows the links of a selected node and thereby reveals the connectivity of that one node. 
    }  
    \label{fig:sampling_types}
\end{figure}

\section*{What is spatial subsampling?}

Depending on the experimental setup, spatial subsampling (see Box 1) can be subject to different constraints that determine which units can be accessed to collect observations. 
These constraints, in turn, affect how subsampling distorts the data. 
        
We will first illustrate three different subsampling approaches using the example of a static, spatially-extended network~\cite{hu_survey_2013,leskovec_sampling_2006} (Fig~\ref{fig:sampling_types}).
In a typical experimental sampling Ansatz, one might observe a random subset of nodes and connections between observed nodes (Fig.~\ref{fig:sampling_types}, lower left panel).
An example of such \emph{random subsampling} is the random selection of members of a social group to construct a contact network~\cite{sekara_fundamental_2016, genois_can_2018}.
While such schemes are designed to sample units in a representative way, it was shown early on that the statistics of the degree distribution is often not preserved under subsampling, but clearly biased~\cite{stumpf_subnets_2005}.
%
A complementary experimental sampling Ansatz is to record a complete set of nodes within a certain observation window (Fig.~\ref{fig:sampling_types}, lower middle).
Such \textit{window sampling} is a common sampling scheme for - in the wider sense - optical-based observations, e.g., when imaging a focus area smaller than the system of interest~\cite{hillman_optical_2007, skocek_high-speed_2018};
or when considering traffic data from a particular urban area~\cite{wang_quasi-stationary_2020}.
The advantage of such a method is the detailed observation of all the connections or interactions in the field of view, the disadvantage is that the sampling window might not be representative, and even if the sampling window is representative, it is subject to finite-size effects.
In some setups, researchers can control the sampling in a snowball-like mode: starting with a selected node, one follows the connections to its nearest neighbours, potentially the next-nearest neighbours, and so on (Fig.~\ref{fig:sampling_types}, lower right).
Such sampling is obtained when using tracer molecules injected in one neuron (or brain area) and observing which neurons are reached by the tracer via physical (dendritic or axonal) connections; likewise, contact tracing in social networks belongs to this class of sampling.
This approach preserves many local properties of the network~\cite{leskovec_sampling_2006, tsitsvero_signals_2016}, however, errors in tracing can impact the results, tracing can be quite laborious and technically intensive~\cite{turner_reconstruction_2022}, and it is not applicable to all systems.  

\begin{figure}
    \centering
\includegraphics[trim=0 0 0 0, clip, width = 0.9 \linewidth]{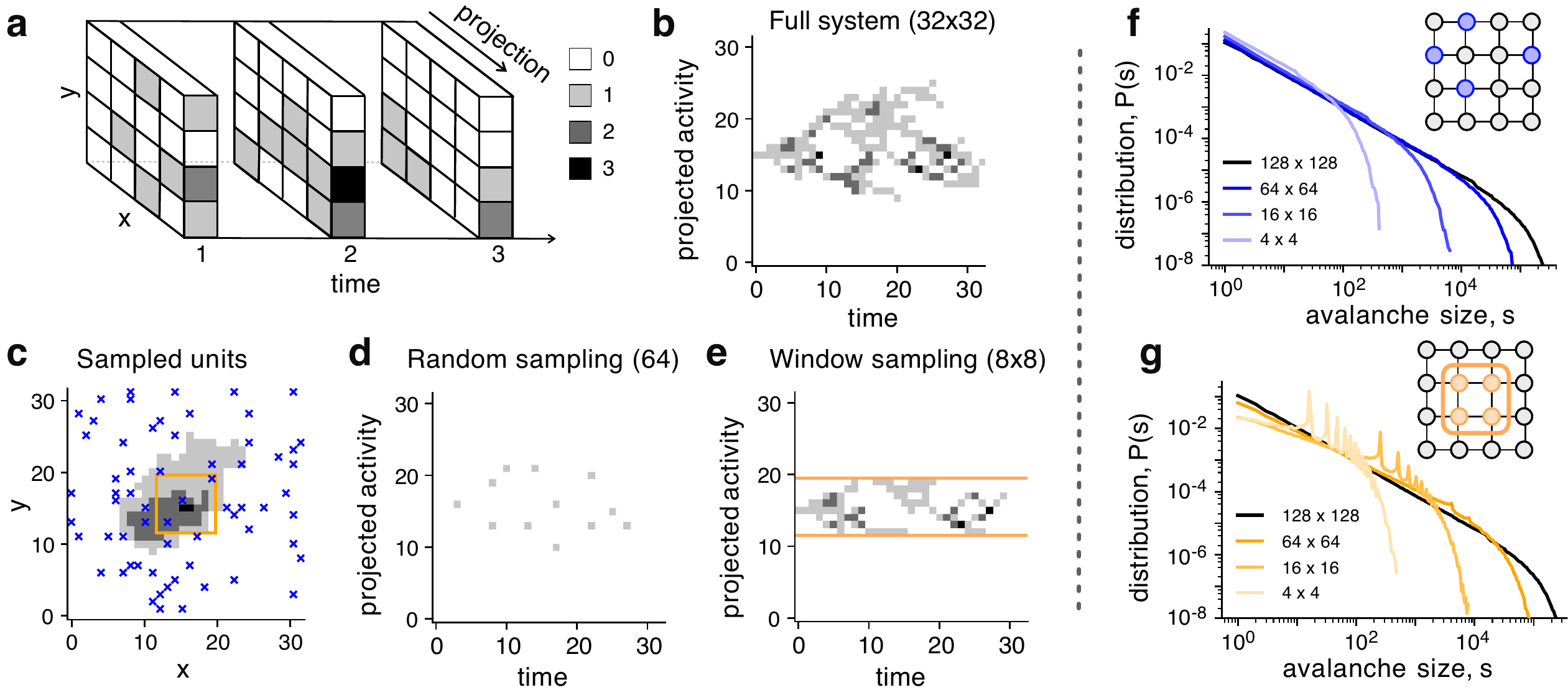}
    \caption{\textbf{Sampling of dynamical observables.} \textbf{a.} Schematics of activity unfolding in the $x$-$y$ plane over time and the corresponding projection onto the $y$ axis. For each time $t$, the color of the column denotes the summed activity, i.e. $a_y(t) = \sum_x a_{xy}(t)$ as indicated in the legend that applies analogously to panels b-e. 
    \textbf{b.} Example of a single activity propagation (avalanche) observed in a complete $32\times32$ Abelian sandpile model~\cite{priesemann_subsampling_2009, bak_self-organized_1987} after a single unit is triggered at time $t=0$. 
    \textbf{c.} Same avalanche but projecting time into the $x$-$y$ plane. 
    \textbf{d.} Same avalanche but from $64$ randomly sampled units (blue crosses in c).
    \textbf{e.} Same avalanche but in a window of $8\times8$ units (orange area in c).
    \textbf{f.,g.} Distribution of avalanche sizes $P(s)$ from a larger  $128 \times 128$ model sampled with different numbers of units $N$ (indicated in the legends). 
    \textbf{f.} For random choice of the $N$ units, the shape of the avalanche-size distribution is approximately (but not perfectly) preserved.
    \textbf{g.} Under window sampling, the distribution of avalanche sizes from the same model can show multiple peaks at $N$, and multiples of $N$. 
    Distributions in f,g were obtained from $10^6$ avalanches on a 2D network for each condition.
    }
    \label{fig:sampling_activity}
\end{figure}

While the study of sampling static graphs and their structure has been advancing in the past decades~\cite{hu_survey_2013,leskovec_sampling_2006}, capturing collective properties of stochastic dynamics in subsampled complex systems is only starting to unfold.
In principle, investigating how the dynamical system unfolds over time provides valuable insights about its dynamics, even if only a fraction of units or dimensions are observed.
For example, the classical Takens or embedding theorem states, that for  deterministic, chaotic systems with a $D$-dimensional attractor, it is sufficient to observe just a single unit (or dimension). Embedding its temporal dynamics based on the past $2D+1$ observations~\cite{rand_detecting_1981}, or even less~\cite{sauer_embedology_1991}, enables one to reconstruct a map of the attractor~\cite{rand_detecting_1981, sauer_attractor_2006, whitney_differentiable_1936, shalizi_homophily_2011}.
With this temporal delay embedding, one can not only predict the evolution of a single observed unit, but one can further infer certain system properties, e.g. Lyapunov exponents, at least approximately. 
While the theorem has been derived for deterministic systems without observational or system noise, delay embedding is also used to study the probabilistic evolution of stochastic systems~\cite{kantz_nonlinear_2004,wibral_bits_2015}.
However, when it comes to inferring collective properties from stochastic dynamics, or full reconstruction of the entire system, one requires more information about the temporal dynamics of sampled units, their correlation and their interaction. 

As for the static networks, the specific subsampling Ansatz can severely affect the observation of collective dynamics from the sampled units.
For example, when sampling activity clusters or cascades such as \emph{avalanches}~\cite{beggs_neuronal_2003,bak_self-organized_1987} from a spatially-structured system (such as the brain), then the observed spatial-temporal activity patterns can strongly deviate from the patterns in the whole system (Fig.~\ref{fig:sampling_activity}a-e)~\cite{priesemann_subsampling_2009,levina_subsampling_2017,wilting_between_2019,ribeiro_undersampled_2014}.  
Distributions of avalanche sizes reveal a statistical difference between sampling types, as can be illustrated using relatively simple models, such as the Abelian sandpile model~\cite{bak_self-organized_1987, dhar_self-organized_1990} on a square lattice (Fig.~\ref{fig:sampling_activity}f-g). 
Under random sampling, the overall shape of the observed data distribution is primarily preserved~\cite{ribeiro_undersampled_2014, levina_subsampling_2017}. 
However, window sampling can generate statistical artifacts that were not present in the original data, as it clearly does not reflect a representative sample.
Here, the avalanche-size distribution can exhibit characteristic peaks if one uses a small observation window~\cite{beggs_neuronal_2003, priesemann_subsampling_2009, chen_avalanche_2011, acebron_kuramoto_2005}. These peaks are not present under random sampling or when sampling the full system; hence they clearly emerge due to the chosen sampling scheme.
The exact impact of sampling strategies on observables is still to be determined. 
In the following, we outline known sampling biases together with routes that enable one to infer collective properties from limited data.

\section*{On solutions for spatial subsampling problems using scaling theory}

Building on the achievements of statistical physics, an intuitive approach to overcome the subsampling problem is to adapt approaches from finite-size scaling theory. Finite-size scaling is a phenomenological theory that has been developed to probe scale invariance in complex systems.
A common theme in such scaling theories is to (i) formulate a multivariate function that describes the dependence of a system property on different parameters, to (ii) identify relevant scales of such parameters, and to (iii) rescale these parameters such that they generate a universal function that no longer depends on microscopic details of the underlying system.
For systems that feature scale invariance, such transformations are expected to be described by so-called critical exponents.
As we demonstrate in the following, such phenomenological scaling theories can naturally be extended to overcome the subsampling problem.


In the following, note that many observables of interest are discrete (e.g., the node degree $k$ is an integer for unweighted networks). This sets an additional minimal scale, and makes claims about scale-free distributions, e.g., $P(k)\sim k^{-\gamma}$, approximate~\cite{clauset_power-law_2009}.
We will nonetheless stick to the common integral description of continuous variables for convenience, but keep in mind that the integrals become sums for discrete variables.

\paragraph*{Phenomenological approach for scale-free degree networks.}
When testing scale invariance of real networks, one is faced with the constraint that real data represents a (potentially small) subsample of an inherently finite network. Both, the finite size of the system, as well as the finite sample thereof, set a natural limit to the maximal extend of collective properties and thereby typically introduce a cutoff, i.e., observing avalanche sizes or node degrees above a characteristic maximal scale $s_0$ becomes very unlikely.

\begin{figure}
    \centering
    \includegraphics[width=0.66\linewidth]{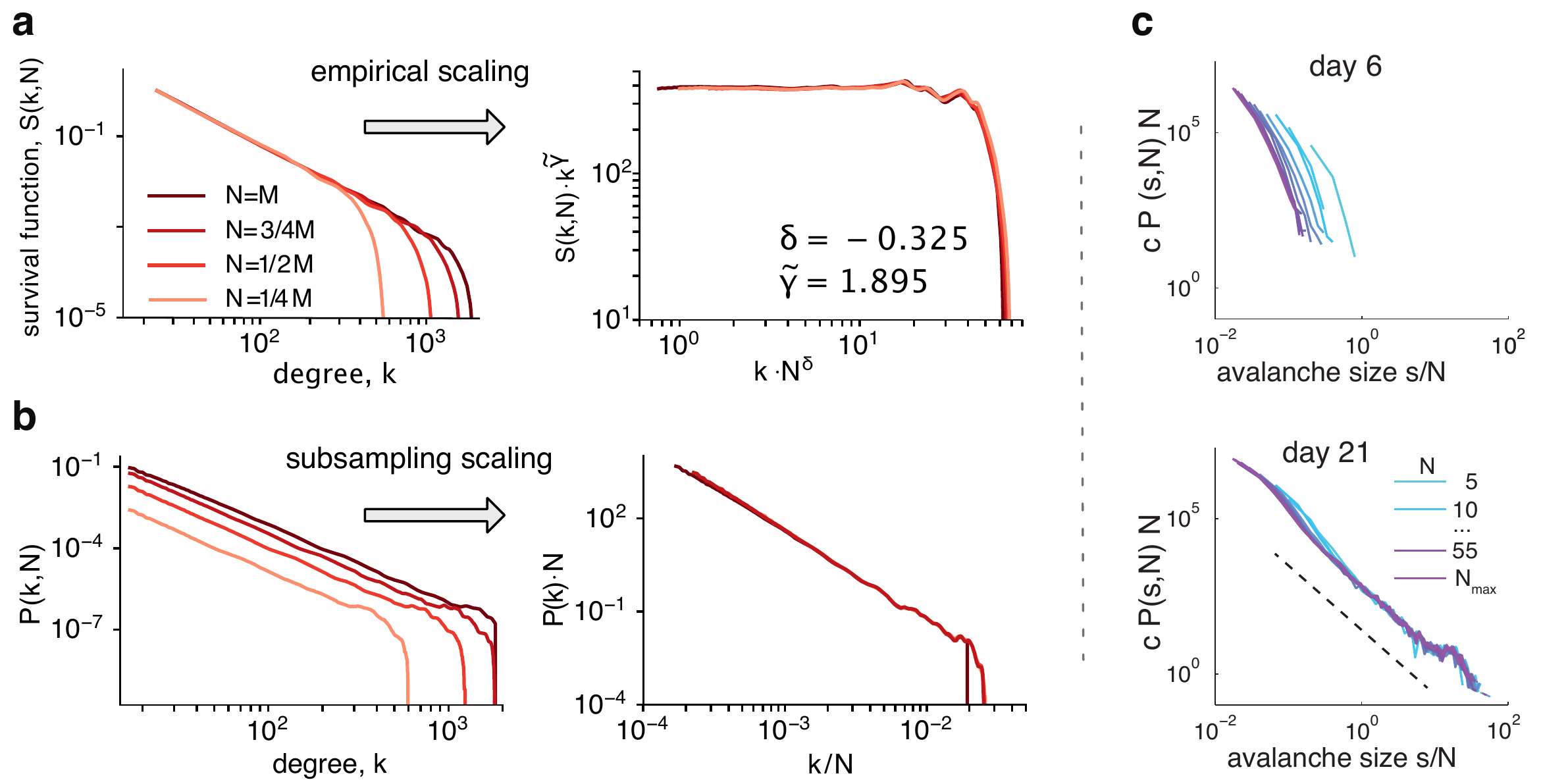}
    \caption{{\bf Scaling solutions for subsampled observations.} 
    \textbf{a.} The survival function $S(k,N)$ of a Barabasi-Albert random network with $M = 10^5$ nodes shows a power-law with finite-size related cutoff. 
    Subsampling is applied by sampling a random set of $N$ network nodes, and keeping only the edges between sampled nodes (subsampled distributions are obtained by averaging 100 resamples). 
    Empirical scaling optimization allows finding parameters to collapse survival functions $S(k,N)$ using numerical fitting routines~\cite{serafino_true_2021}. 
    {\bf b.} Same networks as in a but using the analytically derived ``subsampling scaling'' to collapse the degree distribution $P(k,N)$ without requiring to know or fit the underlying scaling exponent~\cite{levina_subsampling_2017} (the deviation in the cutoff is due to the finite size of the original graph resulting in the hard cutoff in the degree distribution). 
    {\bf c.} Using data from developing neuronal cultures (reproduced from~\cite{levina_subsampling_2017}), one observes that in early stages (top, day 6) the network has not yet developed a scale-free avalanche size distribution $P(s, N)$, cf. Fig.\ref{fig:sampling_activity}, and hence the distributions obtained by further subsampling (color) do not collapse under subsampling; upon maturation the subsampling scaling collapses distributions indicating underlying scale-free avalanche distributions (bottom, day 21). This finding allows to extrapolate the distribution to system size.   
    }
    \label{fig:random_sampling}
\end{figure}

To draw conclusions about collective properties from subsampled systems, one can make use of finite-size scaling tools developed for critical phenomena~\cite{perkovic_disorder-induced_1999, sethna_statistical_2006, chen_avalanche_2011}.
For example, Serafino and colleagues~\cite{serafino_true_2021} proposed a heuristic approach to infer scale-freeness of subsampled networks that operates on the survival function $S(k)$ (i.e., the probability to observe a degree of at least $k$), related to the cumulative distribution function $CDF(k)$ as $S(k)=1-CDF(k)$.
For scale-free networks with a power-law degree distribution $P(k)\propto k^{-\gamma}$, $\gamma>1$, the survival function becomes $S(k) = \int_k^\infty P(q)dq\sim k^{-\tilde{\gamma}}$, with $\tilde{\gamma}=\gamma-1$.
Being an integral over the original degree distribution, the survival function is much smoother than $P(k)$ in the relevant scaling region of large $k$. 
Assuming that finite-size effects lead to a crossover behavior, where for $k<k_0$ the survival function follows a power law, while for $k>k_0$ the distribution falls more rapidly, then one can reformulate the finite-size scaling hypothesis to incorporate finite sample size $N$~\cite{serafino_true_2021}
\begin{equation}
    S(k,N)  = k^{-\tilde{\gamma}}f(kN^\delta),
\end{equation}
with the (unknown) exponents $\tilde{\gamma}>0$ and $\delta<0$, and a universal scaling function $f(\cdot)$. 
Instead of using different system sizes as in common finite-size scaling techniques, one can verify this scaling hypothesis on a single empirical data set by artificial subsampling~\cite{levina_subsampling_2017, serafino_true_2021}. 
For this, one (i) generates multiple subnetworks by randomly selecting $N$ of the original $M$ nodes, (ii) determines $S(k,N)$ for all subnetworks, and (iii) determines the optimal parameters $\tilde{\gamma}$ and $\delta$ by seeking to collapse plots of $S(k,N)k^{\tilde{\gamma}}$ versus $kN^\delta$ (Fig.~\ref{fig:random_sampling}a).
The goodness of the collapse plot provides a measure for the scale-free behavior.
Because of deviations from the scale-free behavior for small degrees, the collapse should only be performed for $k>k_\mathrm{min}$.
This approach revealed that many empirical networks satisfy the finite-size scaling hypothesis of collapse~\cite{serafino_true_2021}, arguing against the observation that scale-free networks are rare~\cite{khanin_how_2006, broido_scale-free_2019} but supporting claims that scale-free properties are a common feature of real complex networks~\cite{clauset_power-law_2009, stumpf_critical_2012, voitalov_scale-free_2019, holme_rare_2019,willinger_more_2004}.

\paragraph*{Subsampling-scaling theory for random subsampling.}
Random subsampling presents a particularly well suited sampling Ansatz for systems where the collective properties of interest are expected to affect every unit.
For random subsampling, one can go beyond heuristic approaches and also beyond scale-free distributions (extending the applicability of the method above), and explicitly calculate scaling transformations that allow to collapse distributions of collective properties from subsampled systems~\cite{levina_subsampling_2017}. 
Such calculations provide exact results for exponential and negative binomial distributions and good approximations for the particularly interesting case of power-law distributions.
In fact, for power laws, the calculations yield precise predictions for universal \emph{subsampling scaling} exponents that are independent of the underlying power-law exponent.

The resulting subsampling-scaling theory can be directly applied to collective properties that follow a power-law distribution $P(x)\sim x^{-\gamma}$, $\gamma > 1$. The power-law distribution can represent static properties, such as degree distributions, or dynamic ones, such as avalanche size distributions.
For power-law distributions, it can be shown mathematically that when sampling $N$ units randomly, the resulting distribution of a collective property $x$ is described by a universal scaling function: 
\begin{equation}
    P(x,N) = N^{-\alpha}f(x \cdot N^{-\beta}), \; \mathrm{for \; any}\; \alpha,\beta \in \mathbb{R} \;\: \mathrm{with}\: \alpha - \gamma \beta = 1 - \gamma. \label{eq:subsampling_scaling}
\end{equation}
Here, the quantity $x$ can be any collective property computed for the observed part of the system, including the degree of a network $x=k$ (cf. comparison to heuristic scaling in Fig.~\ref{fig:random_sampling}b) or the size of an avalanche $x=s$ (Fig.~\ref{fig:random_sampling}c).
The most convenient possibility is to select a solution $\alpha =  \beta = 1 $ that does not require any knowledge about the underlying scaling exponent $\gamma$, and it can be further shown that the scaling function $f(x / N)=P(x,N)N$ takes the shape of the asymptotic distribution with the correct exponent $\gamma$. This solution for subsampling scaling allows one to extrapolate to larger sampling sizes. This is relevant if one applies artificial subsampling to an already subsampled system - and wants to extrapolate to the full system. 

The finite size of real world systems typically leads to a finite-size related cutoff in the power-law, thus properties $x$ larger than a cutoff size $x_\mathrm{cutoff}$ are observed more rarely. 
The scaling of the cutoff location with the number of sampled units is expected to be linear ($x_\mathrm{cutoff}(N) \sim N^{-\eta}$, with $\eta =1 $), and thus the subsampling scaling introduced above with $\alpha =  \beta = 1 $ will collapse the cutoff onsets. 
If $\eta \neq 1 $ the scaling should still be recovered by taking $\beta = \eta$ and $\alpha = 1 + (\beta -1) \gamma$.

One particular application is to evaluate whether the collective dynamics of dissociated neuronal networks \textit{in vitro} during development develops scale-free avalanches (Fig.~\ref{fig:random_sampling}c). However, from the ten-thousands of neurons, only about 60 could be recorded. Here, subsampling scaling of neuronal avalanche size distributions reveals that the  data collapse into power-law distributions with maturation of the neuronal networks. 
In more detail, within three weeks of maturation, the neuronal networks develop towards exhibiting more and more scale-free avalanche dynamics that cover up to four orders of magnitude. This indicates that these neural networks self-organize towards the vicinity of a critical point~\cite{beggs_neuronal_2003, girardi-schappo_critical_2013, munoz_colloquium_2018,wilting_operating_2018,girardi-schappo_synaptic_2020,neto_unified_2020}.

\paragraph*{Scaling analyses for window subsampling.}
When using window sampling, the correlations emerging from local interactions may cause systematic deviations in observed collective quantities~\cite{magni_visualization_2009, chen_avalanche_2011, priesemann_subsampling_2009}.
We showed in Fig.~\ref{fig:sampling_activity}g
an example of avalanche-size distributions from a recording window that is considerably smaller than the system size~\cite{priesemann_subsampling_2009} 
Applying window sampling to the classical Bak-Tang-Wiesenfeld model~\cite{bak_self-organized_1987,priesemann_subsampling_2009} leads to very characteristic effects because the avalanches are proven to be spatially compact~\cite{dhar_abelian_1999}.
Thus, in a compact observation window, there is an increased probability that the avalanche will activate all observed units, and hence the probability to observe avalanches of the size of the observation window - or multiples thereof - is increased. 
That leads to the multi-peaked distribution~(Fig.~\ref{fig:sampling_activity}g). Moreover, even beyond these peaks the slope of the power-law distribution is altered. 
This example may be extreme, but it illustrates how spatial sampling features (i.e., sampling data from a compact window)  can strongly distort the observables, and that fitting avalanche distributions may then turns into a multivariable scaling analysis with two or more control parameters that requires systematic analyses e.g. via Bayesian inference~\cite{chen_avalanche_2011}.

Finite observation windows, however, have the advantage that they directly provide detailed spatial information within each window. In many systems one can thus again adapt finite-size scaling approaches to extrapolate to the full system from artificial subsampling.
For example, Martin and coworkers recently proposed the so-called \textit{box scaling} procedure~\cite{martin_box_2021} that considers the instantaneous connected correlation function $C(r)$ as a suitable indicator to distinguish critical (dynamical) states from non-critical ones. $C(r)$ measures the average correlation of units separated by a distance $r$ for a specific snapshot in time. 
Mathematically, this can be expressed as 
\begin{equation}
    C(r,t) = \frac{1}{c_0}\frac{\sum_{i,j} \left(s_i(t)-\overline{s}(t)\right)\left(s_j(t)-\overline{s}(t)\right)\delta(r_{i,j}-r)}
    {\sum_{i,j}\delta(r_{i,j}-r)},
\end{equation}
where the ``smoothed'' Dirac-$\delta$-function $\delta(r_{i,j}-r)$ considers pairs with approximate distance $r_{i,j}\approx r$ and the normalization constant $c_0$ ensures that $C(0,t)=1$.
Note that this definition only considers the correlation with respect to the instantaneous mean $\overline{s}(t)$, which neglects temporal correlations of the global system but is argued to provide an estimate of the spatial correlation length $r^\ast$ from the zero-crossing of the correlation function~\cite{martin_box_2021, cavagna_scale-free_2010, grigera_correlation_2021}.

\begin{figure}
    \centering
    \includegraphics[trim=0 0 0 0, clip, width = 0.63\linewidth]{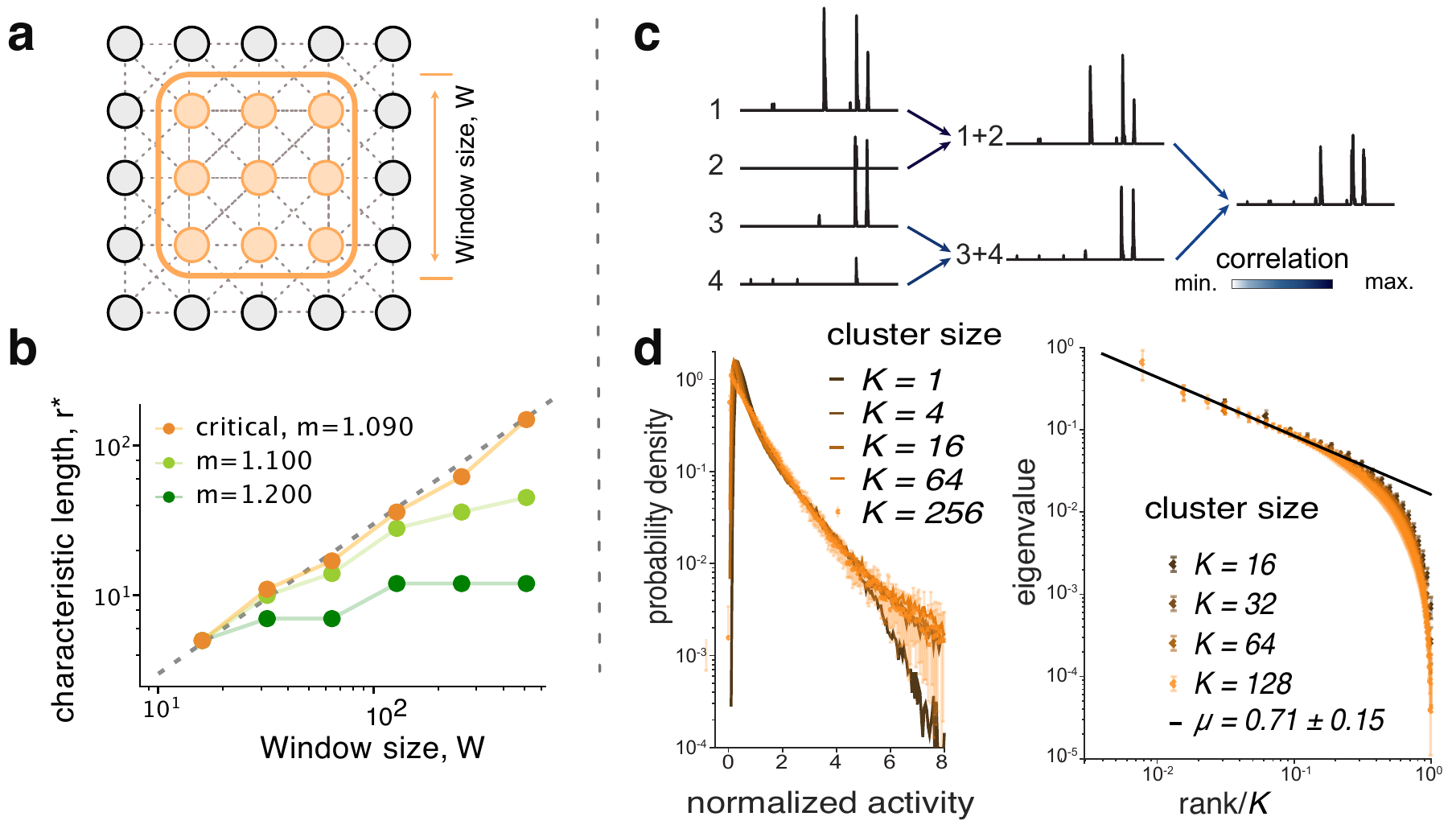}
    \caption{\textbf{Scaling theory can reveal macroscopic properties of collective dynamics despite window sampling.} {\bf a.} Schematics of window sampling in a locally connected two-dimensional network, where active units can activate each of their $8$ nearest neighbors with probability $m/8$. 
    {\bf b.} From the scaling of the spatial correlation length $r^\ast$ in artificially subsampled windows of size $W$, one can assess whether the population dynamic of the system is close-to-critical or not~\cite{martin_box_2021}.
    {\bf c.} A complementary, heuristic coarsening scheme has recently proposed for neuronal activity where the underlying connections are unknown.
    Here, one first determines a hierarchy of correlations between pairs of neuronal activities, to then group maximally correlated pairs of neurons by summing their activities, and normalizing such that the mean of non-zero values is one. 
    This summation procedure is repeated $k$ times to obtain neural clusters of size $K=2^k$.
    {\bf d.} Scaling of the normalized activity and eigenvalues for different cluster sizes $K$ reveals signatures of critical behavior. Panels (c) and (d) reproduced from~\cite{meshulam_coarse_2019}.}
    \label{fig:box_scaling}
\end{figure}

From the scaling of $r^\ast$ in artificially subsampled observation windows (or box scaling), one can distinguish critical from non-critical systems.  
At a critical point, where the asymptotic correlation length $\xi$ diverges, $r^\ast$ is bounded by the linear size $W$ of an observation window such that $r^\ast\sim W$ (when $\xi\gg L\gg W$).
Away from the critical point, $\xi$ approaches a finite value and one can show that $r^\ast\sim \xi\log(W/\xi)$ (when $\xi\ll W<L$).
Indeed, this approach can be illustrated for model systems on 2-dimensional lattices (Fig.~\ref{fig:box_scaling}a,b).
Moreover, this formalism enables to study how well the correlation function can be collapsed under rescaling, showing good data collapse for critical dynamics but no collapse for non-critical dynamics, signalling the absence of scale invariance captured in a single scaling function.
Applied to human brain activity from fMRI data~\cite{fraiman_what_2012,tagliazucchi_criticality_2012} this method confirms linear scaling~\cite{martin_box_2021} compatible with neural dynamics showing scale-free properties~\cite{beggs_neuronal_2003, girardi-schappo_critical_2013, munoz_colloquium_2018, carvalho_subsampled_2021}.

\paragraph*{Phenomenological renormalization group.}
Complementary statistical-physics inspired approaches to study scale invariance from subsampled activity involve phenomenological coarse-graining procedures motivated by the renormalization group. 
However, in many systems the detailed neighbourhood information for the coarse-graining is inaccessible, requiring empirical approaches.
For example, Meshulam and colleagues~\cite{meshulam_coarse_2019} recently proposed a heuristic coarse-graining scheme to compute the pairwise correlation matrix of all state variables, sort pairs according to their correlation, and then coarse-grain these pairs by summing their states (Fig.~\ref{fig:box_scaling}c).
Iterative application of this scheme yields distributions of coarse-grained variables that approach a fixed non-Gaussian form indicating evidence of scaling in both static and dynamic quantities (Fig.~\ref{fig:box_scaling}d).
In more detail, coarse-graining data into clusters of size $K$, they showed that the distribution of non-zero activity approaches a fixed non-Gaussian form and that the eigenvalues $\lambda$ of the covariance matrix can be described by the scaling function $\lambda \propto (K/\mathrm{rank})^\mu$ with $\mu\approx 0.7$~\cite{meshulam_coarse_2019}, without identifying the 
 specific universality class for the neural data.
Coarse-graining data from model simulations, where the distance to the critical point can be controlled, it was shown that such phenomenological coarse-graining procedures indeed have the potential to distinguish critical from non-critical systems~\cite{nicoletti_scaling_2020}. 
However, these results remain controversial and they might not be specific to critical systems only: First, some scaling features apparently persist also in the supercritical regime for finite systems, rendering a distinction difficult~\cite{nicoletti_scaling_2020}.
Second, similar scaling features as well as the flow towards a non-Gaussian fixed point can be explained by the presence of multiple unknown, time-varying latent fields to otherwise independent units~\cite{morrell_latent_2021}.
Third, using real-space coarse-graining schemes with continuous sampling units (e.g., in measurements with local field potentials), further runs the risk to introduce spurious correlations into the measurement that can strongly affect estimates of collective phenomena such as avalanche size distributions in neural networks or total magnetization in the Ising model~\cite{neto_unified_2020}.
Coarse-graining procedures thus have strong potential to study scale invariance of systems only accessible through finite observation windows but great care has to be taken as hidden latent fields or even the sampling process proper can make a system appear scale free.



\section*{On solutions for spatial subsampling problems beyond scaling theory}  


While in the past sections, we investigated universal scaling behavior, here we turn to suitable subsampling-invariant observables and inference approaches that allow overcoming the effects of unobserved units.

\paragraph{Inferring collective dynamics from stationary subsampled activity.}
\begin{figure}
    \centering
    \includegraphics[width=0.76 \linewidth]{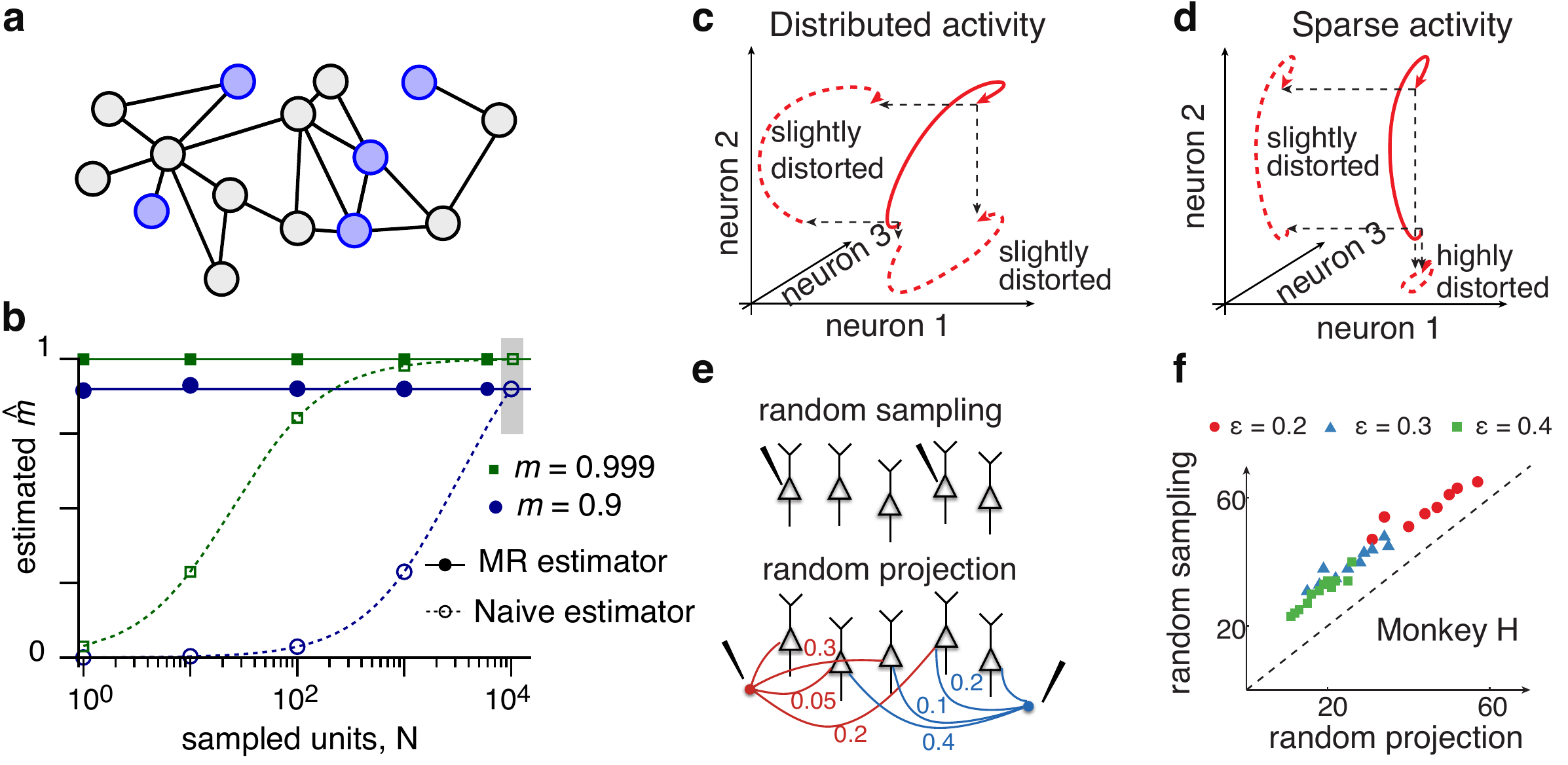}
    \caption{{\bf Subsampling-invariant measures.} {\bf a.} Sampling neuronal activity from a small part of the network (blue units only). 
    {\bf b.} Running a spreading process on a random graph, a na\"ive estimator of the branching parameter $m$ (or equivalently the reproduction number $R$) returns biased results for sample sizes $N$ smaller than the full network ($M=10.000$, gray shading; dashed lines theory, empty symbols numerical simulations). 
    However, the subsampling-invariant ``MR-estimator'' recovers the unbiased branching parameter $m$ (full lines). For many recurrent networks, one can infer $m$ even when sampling the activity of a single unit only ($N=1$)~\cite{wilting_inferring_2018}. 
    Results are illustrated for two different branching parameters, $m=0.9$ and $m = 0.999$.
    {\bf c-f.}
    In noisy systems, the amount of information that is preserved in a subset of neurons depends on the statistics of the population activity: {\bf c.} For distributed activity, a lot of the dynamic range is preserved, and the subsampled activity typically is only slightly distorted, {\bf d.} while for sparse activity different subsets can yield very distorted representations, making inference about the full system challenging.
    {\bf e.} 
    To estimate a minimal number of randomly subsampled neurons required to achieve a representative sample, one can build on  a theory developed for random projections. 
    {\bf f.} Recordings from monkey premotor cortex support the hypothesis that geometric distortions from random projections and random sampling are correlated. $\varepsilon$ denotes the levels of distortion. Panels c-f taken with permission from Ref.~\cite{gao_theory_2017}.}
    \label{fig:invariant_measures}
\end{figure}

The population dynamics of many complex systems, such as the spread of a virus in a population or the spread of neural activity in a network, can be described by a branching process, which is a stochastic process with an autoregressive representation~\cite{harris_theory_1963, kersting_unifying_2020}. 
In a branching process, the average activity (number of infected individuals or active neurons at time step $t$) depends only on the activity in the previous time step or generation:
\begin{equation}\label{eq:mr}
    \langle A_{t+1}|A_t\rangle = m A_t + H,
\end{equation}
where $A_t$ is a discrete number of active units (infected people or spiking neurons at generation $t$), $m$ is the branching parameter (equivalent to the reproductive number $R$ in disease spread), and $H$ is the mean external input. Inherently, the process is stochastic, because at every time-step, the number of offsprings (e.g. the number of new infections $Y$ caused by each infectious person $i$ in the previous time-step) is a random variable. 
The \textit{mean} number of offsprings $m=\langle Y \rangle$ determines whether one observes on expectation a growth ($m>1$), a decline ($m<1, H=0$), a steady state ($m<1, H>0$), or critical phenomena ($m=1$, $H\to 0$). 
Inferring the branching parameter or reproductive number from observed activity is important to understand the underlying spreading dynamics. 

In many complex systems, one needs to infer $m$ from possibly heavily subsampled observation $\tilde{A}_t$ of the total activity $A_t$.
A classical approach to estimate the branching parameter would be solving Eq.~(\ref{eq:mr}) for observed data using linear regression, i.e., $r_1=\mathrm{Cov}\left[\tilde{A}_{t+1},\tilde{A}_t\right]/\mathrm{Var}\left[\tilde{A}_t\right]$~\cite{heyde_estimation_1972,wei_estimation_1990}.
This classical estimator, however, is severely biased when applied to a subsampled population due to the intrinsic noise of the units or the sampling process (Fig.~\ref{fig:invariant_measures}).
Instead, it was proposed to use a multi-step regression (MR) estimator~\cite{wilting_inferring_2018,wilting_operating_2018}, which allows to infer the correct branching parameter $m$ even from heavily subsampled data.
The mathematical basis for this methods is to infer an autocorrelation timescale $\tilde{\tau} = - \Delta t/\ln(m)$ of the process $\tilde{A}_t$ that for random spatial sampling can be shown to remain invariant under subsampling. Thus, while the absolute values of the autocorrelation strength are biased by subsampling, the autocorrelation time is invariant.
This estimator was later generalized to cyclostationary processes and to repeated non-stationary input~\cite{de_heuvel_characterizing_2020}, a common situation for seasonality in disease spread or trial-based neuroscience experiments, respectively (see also Ref~\cite{spitzner_mr_2021} for a toolbox implementation). 
A recently developed Bayesian method allows to increase the accuracy of timescales estimation and thereby enables inference and model selection of cases with various combinations of timescales and oscillation~\cite{zeraati_flexible_2022}.  

The idea to characterize the state of a complex system by the so-called intrinsic timescale was explored even before it was shown to be invariant under subsampling.
Prominently, strongly subsampled spike recordings from different brain areas revealed that intrinsic timescales reflect the functional hierarchy in the neocortex~\cite{murray_hierarchy_2014, siegle_survey_2021} and areal differences in temporal integration of information~\cite{hasson_hierarchical_2015, raut_hierarchical_2020, spitmaan_multiple_2020, gao_neuronal_2020}. 
The multi-step regression  was successfully applied to investigate memory and learning in neuronal networks~\cite{skilling_critical_2019}. 
In models, it was shown that longer intrinsic timescales (and $m$ close to unity) improves performance on tasks that require temporal integration; however, for tasks that are less complex, a smaller branching parameter $m$ led to better performance~\cite{cramer_control_2020}. 
In that light, the longer intrinsic timescales observed in experiments for higher brain areas indeed may point to longer and more complex integration of information.

A precise estimate of the strength of recurrent interactions helps to shed light on neural processing. 
If recurrent interactions are classified with the branching parameter, one finds that for close to $m=1$, small changes in the recurrent (synaptic) coupling strength between neurons (e.g. via neuromodulation) can alter the intrinsic timescale and thereby the processing properties of the network considerably~\cite{wilting_operating_2018, cramer_control_2020}. 
A complementary approach to estimate the strength of recurrent interaction in terms of the recurrency $R$ is based on the cross-correlation distribution across neurons~\cite{dahmen_second_2019}, and argues along the same line: With recurrency close to $R=1$ networks are in a sensitive regime where small changes in the coupling statistics lead to considerable changes of network properties, including the spatial pattern of neuronal coordination~\cite{dahmen_global_2022} and the effective dimensionality of network activity~\cite{dahmen_strong_2022}.
The sensitivity to small changes in the coupling arises because in the vicinity of non-equilibrium phase transitions (e.g. for branching network $m=1$), a number of properties like the intrinsic timescale diverge. Therefore, understanding how $m$ or complementary the recurrency or dimensionality change across brain areas and with task conditions, will shed light on the self-organization of neural processing in vivo.

Several recent experiments study how the branching parameter $m$, the recurrency and the intrinsic timescale change with task conditions. First, it was shown that the intrinsic timescales of local activity in a brain area increases with attention~\cite{zeraati_attentional_2021-1}, indicating a shift of the neural network dynamics towards the critical point. Second, after a perturbation by stopping sensory input, neural networks transiently decrease their branching parameter $m$ but then recover a value close to unity, and thus longer intrinsic timescales, within few days, presumably via homeostasis~\cite{hengen_neuronal_2016}. Third, cortical stimulation is detected better, when the network shows lower instantaneous recurrency before the stimulus is applied, indicating that the larger signal-to-noise ratio may facilitate detection of such stimuli~\cite{rowland_perception_2021}. 
All of these results in neuroscience explicitly or implicitly rely on the fact that autocorrelation timescales or estimated recurrency are invariant under observation of a small part of the system.

\paragraph{Quantifying dimensionality of population activity.}
Even when sampling only a small fraction of a complex system, the signals of individual units can be strongly correlated due to the coupling with a system that imposes coordinated population dynamics. 
This property was characterized recently for brain activity, where even most advanced experiments sample only hundreds or thousands out of the many millions of behaviorally relevant neurons.
For that sampled activity, dimensionality reduction methods revealed a striking simplicity underlying multivariate neuronal data recorded during execution of different tasks, by various species from different brain areas \cite{cunningham_dimensionality_2014}, namely that high-dimensional neural trajectories for specific computations resided in much lower-dimensional manifolds~\cite{gallego_neural_2017, gallego_cortical_2018, stringer_high-dimensional_2019, feulner_neural_2021}.
This naturally raises questions, whether the inferred representation would change if more neurons were recorded and consequently how reliable the conclusions are.
Recently, Gao and colleagues derived that a sufficient condition to guarantee a representative manifold is that the number of recorded neurons exceeds a so-called neural task complexity (NTC)~\cite{gao_theory_2017}.
In more detail, for a task with $K$ parameters (e.g., time), the neural task complexity is proportional to the product of  parameter-specific ranges $L_k, \, k=1,\ldots, K$ (e.g., task duration) and inversely proportional to the product of parameter-specific correlation lengths $\lambda_k$ (e.g., population autocorrelation time, for estimation methods see above), i.e.,
\begin{equation}
    \mathrm{NTC} \propto \frac{\prod_k L_k}{\prod_k \lambda_k},
\end{equation}
provided that the unrecorded neurons are statistically similar to the recorded ones.
This theory implies a task-dependent lower bound on the necessary number of random projections (the dimensionality) to accurately recover the geometry of the underlying neural manifold.
Moreover, it was shown that for sufficiently distributed activity (see Fig.~\ref{fig:invariant_measures}c vs. Fig.~\ref{fig:invariant_measures}d) random projections can be replaced with random subsamples of neurons (Fig.~\ref{fig:invariant_measures}e-f) that are experimentally accessible~\cite{gao_theory_2017}.
Importantly, the geometric distortion (the largest relative difference between the distances of neural activity patterns in projected and full space) grows with the logarithm of the number of neurons in a circuit, such that a modest number of neurons (which though increases with task complexity) should suffice for reliable statements about the geometry of neural manifolds.

\paragraph{Inverse problems on subsampled data.}
\begin{figure}
    \centering
    \includegraphics[width=0.96 \linewidth]{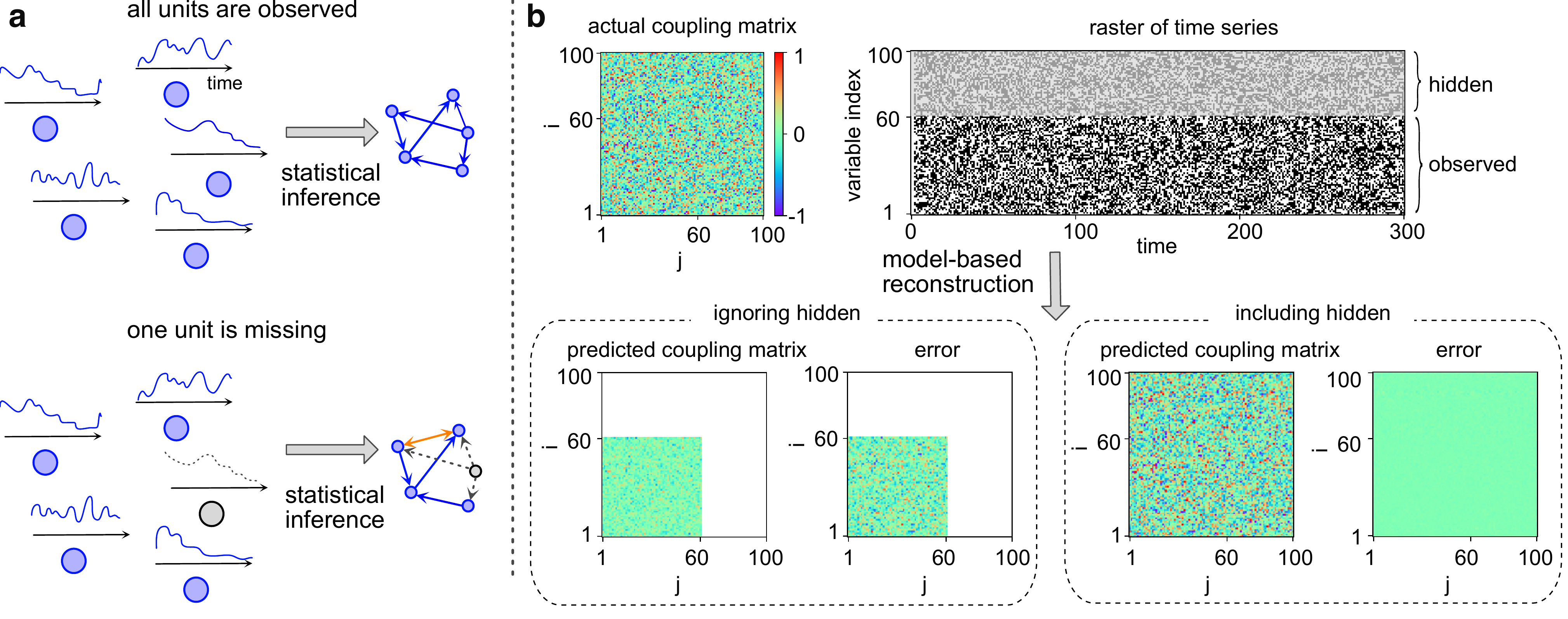}
    \caption{{\bf Reconstruction of a network under subsampling.} {\bf a} Top: Even when observing the whole dynamics of each node, a model-free reconstruction of the dependencies network is a challenging problem that can be, in most cases, solved with various methods of statistical inference. 
     Bottom: Missing observation of nodes dynamics without additional considerations and modeling of confounders often leads to the wrong reconstruction of the dependencies between observed variables (blue - observed nodes and dynamical traces, gray - unobserved, orange - spurious inferred connection).
    {\bf b.}  When assuming a particular dynamical model, reconstruction of the connectivity is possible if it is explicitly introduced into the fitting of the observed units' time series. Panel b is modified from Ref.~\cite{hoang_data-driven_2019}.}
    \label{fig:network_reconstruction}
\end{figure}
The recent explosion in data availability has fostered interest in so-called inverse problems, where the goal is to calculate from a set of observations the causal relations that produce them.
These causal relationships can be implicit, e.g., in the form of hidden latent variable  models~\cite{yu_extracting_2006, macke_empirical_2011, cunningham_dimensionality_2014}, or explicit, e.g., reconstructing networks from incomplete data using statistical (possibly Bayesian or causal) inference~\cite{butts_network_2003,newman_network_2018, young_bayesian_2021}.
Even if there is direct information about a set of nodes and only the edges are uncertain, network reconstruction can still present a non-trivial problem, especially in the presence of higher-order interactions~\cite{battiston_physics_2021}. 
To reduce inference complexity, one can, for instance, consider explicit generative network models~\cite{peixoto_reconstructing_2018, peixoto_network_2019} or focus on properties of interest such as community detection~\cite{hoffmann_community_2020}. 
These methods can be efficient in inferring the correct network from noisy network measurements; however, if one has no direct information about the edges (as typical in biological networks), one requires alternative approaches purely based on node dynamics.


Often the network has to be reconstructed from observed node dynamics without knowing the mechanistic model that gives rise to it (Fig.~\ref{fig:network_reconstruction}a). Network reconstruction, however, belongs to the class of NP-hard problems and thus poses a major computational challenge~\cite{welch_algorithmic_1982}. Moreover, in general correlation and causation are hard to disentangle without causal interventions~\cite{shalizi_homophily_2011,ay_information_2008,peters_elements_2017}.
Practically, the simplest approach is to consider nodes in pairs (or sometimes triplets) and estimate dependencies between them with model-free statistical inference, e.g., using correlation-based measures~\cite{rubinov_complex_2010}, Granger causality~\cite{seth_granger_2015}, or various information-theoretical measures~\cite{stetter_model-free_2012, novelli_large-scale_2019, wollstadt_graph_2015, wibral_bits_2015}.
For these methods, hidden (unobserved) nodes can be considered as latent confounders that induce spurious correlations --- and hence links --- where there are none~\cite{ramb_impact_2013, geiger_causal_2015, runge_causal_2018} (Fig.~\ref{fig:network_reconstruction}~bottom).
To treat confounders, and thereby also unobserved nodes, there have been efforts to include causality in network reconstruction~\cite{spirtes_causation_2000, elsegai_network_2015, shiells_effect_2017, williams-garcia_unveiling_2017, runge_inferring_2019}.
In addition, one can include higher-order correlations between nodes~\cite{schneidman_network_2003} by capturing the statistics of observed patterns in models with pairwise interactions~\cite{schneidman_weak_2006,tkacik_searching_2014}.
For binary observations (such as the presence or absence of spike),this results in the class of inverse Ising problems~\cite{roudi_mean_2011,nguyen_inverse_2017}, which have been applied to many systems, including neural networks~\cite{schneidman_weak_2006} and gene-regulatory networks~\cite{lezon_using_2006}.
In fact, the inverse Ising model can be also formulated with higher-order interactions~\cite{wang_full_2022}.
Moreover, it can be extended with effective non-equilibrium dynamics~\cite{nguyen_inverse_2017, mezard_exact_2011, zeng_maximum_2013, timme_revealing_2014, bachschmid-romano_variational_2016}.
Combining such kinetic inverse Ising models with approximate Bayesian inference was able to capture low-order statistics despite sampling only a fraction of network nodes~\cite{donner_inverse_2017}.
This shows the breadth of approaches for network reconstruction from node dynamics when the underlying dynamics are unknown.
How reliable these reconstructions are and how well they capture the effect of hidden nodes depends on how challenging the problem is and how good the data are.
If one can approximate the mechanistic model giving rise to node dynamics, this opens several opportunities to capture the effect of hidden nodes.
For example, if only a few nodes are absent then this likely affects only the reconstruction of their local neighborhood, causing unstable results with abnormally dense connections, which can be exploited to identify hidden nodes~\cite{su_detecting_2012, su_uncovering_2014}.
A more systematic way is to incorporate the effect of unobserved nodes in an effective description of the observed dynamics~\cite{chorin_optimal_2000, rubin_memory_2014, bravi_statistical_2017}:
In an illustrative manner, the effective dynamics of a subsampled node $x_i$ can be written in the form 
$
    \dot{x}_i(t) = \sum_j W_{ij} x_j(t) + \sum_j \int_0^t dt^\prime M_{ij}(t-t^\prime)x_j(t^\prime) + \chi_i(t)$,
where the first sum describes the local subnetwork dynamics with interactions $W_{ij}$, and the other terms incorporate the effect of unobserved nodes as (non-trivial) memory $M_{ij}(\tau)$ and noise $\chi_i(t)$ terms.
This is similar in spirit to delay embedding or Takens theorem~\cite{rand_detecting_1981, sauer_attractor_2006, whitney_differentiable_1936, shalizi_homophily_2011}.
Including these memory and noise terms when analyzing subsampled data can improve quantitative assessments of collective properties, for example, when applied to identify dominant memory channels in the signalling network of epidermal growth factor receptor~\cite{rubin_memory_2014}, or to distinguishing sources of noise in protein interaction~\cite{bravi_systematic_2020} and gene regulation~\cite{herrera-delgado_memory_2018, herrera-delgado_tractable_2020} networks.
In addition to indirect approaches, one can explicitly include the hidden states in Expectation-Maximization-like algorithms~\cite{dunn_learning_2013, bachschmid-romano_inferring_2014}. 
The basic idea here to split the inverse problem into an iterative 2-step procedure, where one step is to infer the weight matrix based on all measured (observed) and estimated (hidden) states, and another step is to update the hidden states based on the observed states and the weight matrix.
Such an iterative scheme can, in principle, reconstruct network structure and dynamics from subsampled systems (Fig.~\ref{fig:network_reconstruction}b). Example applications range from subsampled data of neural dynamics in salamander retina, reconstructing from selected input neurons the activities of the remaining ones, and of financial dynamics in the S\&P 500 stock market, improving predicted price changes relevant for profitable trades~\cite{hoang_data-driven_2019}.



\section*{Open questions in the field \& outlook}

We presented an overview of promising new techniques to study collective properties of complex systems from partial observations. 
The development of these new approaches is largely driven  by the  rapid development of  experimental techniques that allow increasingly better - but yet far from complete - access to natural complex systems.
Does this development mean that we can simply lean back and wait until sampling techniques are sufficiently good for a given analysis methods?
Certainly not!
Despite ongoing progress in experimental techniques, there probably will always be systems that cannot be sampled fully for technical reasons. 
For large neural systems, like mammalian brains, for example, it is unclear how one would ever record the activity of all neurons in the deeper brain areas in parallel, without harming brain tissue.
It thus remains of ongoing interest to complement existing analysis methods to learn most from subsampled data.

Understanding complex systems can strongly profit from combining observations on different scales.
The integrated evaluation of multi-scale measurements - with different limitations and distortions at each scale - does present a formidable challenge, but it promises to overcome the limitations of sampling at each scale in a synergistic manner, and thereby strongly improve inference about the multiscale activity of the full system. 
To exemplify the recording technique, data from neuronal cultures or slices, or even \emph{in vivo} recordings~\cite{qiang_transparent_2018} can include simultaneous recordings using imaging techniques of the calcium fluorescence or electrophysiological recordings from a multi-electrode array.  
Each recording technique has its weaknesses and advantages: Ca imaging lacks temporal resolution and requires filtering to extract the activity individual cells, but has a large spatial yield; complementary, multi-electrode recordings only access a tiny fraction of the neurons, but record the activity at full temporal resolution.
A promising new avenue of research is thus to improve the mathematical toolset that combines data from such imperfect but complementary recording techniques. 
On a theoretical level, developing a unified theory for  window and random subsampling might require a new scaling theory but if successful could greatly forward an accurate assessment of a systems behavior. 
Therefore, considering the combination of different sampling techniques, also across scales, appears to be highly promising, as the interesting emergent dynamics typically evolves across multiple scales. 

An important, but open question for future research is to better understand how to combine samples across temporal scales. 
Especially for living systems, which are continuously changing, this is by no means a trivial problem. 
Sticking with the example of the brain, it was shown by monitoring the same neurons over time for a repeated task that neural representations may change spontaneously over time~\cite{ziv_long-term_2013,hainmueller_parallel_2018, gonzalez_persistence_2019}.
However, in many systems experimental constraints might not allow to access multiple spatial scales at the same time but only successively. That might hinder an understanding how interactions at the microscopic scale impact emergent behavior at the macroscopic scale and vice versa. 
Integrating information accessed at different spatial scales, even if not recorded in parallel, is a further open challenge, however, we expect important contributions in the field of non-equilibrium physics, to characterize the effect of subsamples across temporal and spatial scales in well-controlled but non-stationary systems.

A gold standard in studying experimental systems are the application of causal interventions~\cite{pearl_causality_2009, rowland_perception_2021}. Combining causal interventions at the microscopic scale with recordings at multiple scales may shed light on their interaction mechanisms. Designing optimal protocols for causal interventions under a given subsampling scheme might further accelerate the understanding of emergent phenomena in complex systems across scales.

Overall, the recent development of sampling techniques together with systematic analysis approaches promises that spatial subsampling theories will become more extensive and, with more extensive data availability, also more powerful.
This invites future research to tackle the next set of questions, in particular designing innovative sampling strategies (how to select the sampled components), how to deal with non-representative samples in general, how to combine spatial subsamples of a different kind (e.g., multi-scale sampling), and how to design causal interventions.
Developing the mathematical tools to tackle these questions promises to open the path to novel, unbiased insights into complex systems and collective dynamics.

\section{Competing interests}
The authors declare no competing interests.

\section*{Acknowledgements}
We would like to thank David Dahmen, Moritz Helias, Malte Henkel, and Peter Sollich for helpful discussions.

\bibliography{2021_Subsampling}

\end{document}